\documentclass[twocolumn,showpacs,nofootinbib,preprintnumbers,amsmath,amssymb,showkeys,superscriptaddress]{revtex4-1}
\usepackage{graphicx}
\usepackage{amsmath}
\usepackage{amssymb}
\usepackage{dcolumn}
\usepackage{bm}
\usepackage{color}
\usepackage{dsfont}
\usepackage{epsfig}
\usepackage{setspace}

\newcommand \beq{\begin{eqnarray}}
\newcommand \eeq{\end{eqnarray}}

\newcommand\eqn[1]{(\ref{#1})}      
\newcommand\Eqn[1]{Eq.~(\ref{#1})}  
\newcommand\Fig[1]{Fig.~\ref{#1}}  

\begin{document}
\allowdisplaybreaks

\title{Universal aspects of the phase diagram of QCD with heavy quarks}

	\author{J. Maelger}
	\affiliation{Centre de Physique Th\'eorique, Ecole Polytechnique, CNRS, Universit\'e Paris-Saclay, F-91128 Palaiseau, France.\vspace{.1cm}}
	\affiliation{Astro-Particule et Cosmologie (APC), CNRS UMR 7164, Universit\'e Paris Diderot,\\ 10, rue Alice Domon et L\'eonie Duquet, 75205 Paris Cedex 13, France.\vspace{.1cm}}
	
	\author{U. Reinosa}%
	\affiliation{%
		Centre de Physique Th\'eorique, Ecole Polytechnique, CNRS, Universit\'e Paris-Saclay, F-91128 Palaiseau, France.\vspace{.1cm}}%
	
	\author{J. Serreau}
	\affiliation{Astro-Particule et Cosmologie (APC), CNRS UMR 7164, Universit\'e Paris Diderot,\\ 10, rue Alice Domon et L\'eonie Duquet, 75205 Paris Cedex 13, France.\vspace{.1cm}}

	\date{\today}

\begin{abstract}
{The } flavor dependence of the QCD phase diagram presents universal properties in the heavy quark limit. For the wide class of models where the quarks are treated at the one-loop level, we {show, for arbitrary chemical potential,} that the flavor dependence of the critical quark masses---for which the confinement-deconfinement transition is second order---is insensitive to the details of the (confining) gluon dynamics and that the critical temperature is constant along the corresponding critical line. We illustrate this with explicit results in various such one-loop models studied in the literature: effective matrix models for the Polyakov loop, the Curci-Ferrari model, and a recently proposed Gribov-Zwanziger-type model. We further observe that the predictions which follow from this one-loop universality property are well satisfied by different calculations beyond one-loop order, including lattice {simulations}. For degenerate quarks, we propose a simple {universal} law for the flavor dependence of the critical mass, {satisfied} by all approaches.

\end{abstract}

\maketitle

Understanding the properties of strongly interacting matter in extreme conditions of temperature, density, magnetic field, etc., is a question of topical interest with various implications, e.g., in early universe cosmology and astrophysics \cite{Weissenborn:2011qu}. This relates to some of the most profound aspects of quantum chromodynamics (QCD), namely, the physics of (de)confinement and of chiral symmetry breaking/restoration. Past, present, and upcoming heavy-ion collision experiments at RHIC, CERN, FAIR, NICA and J-PARC, accompanied by an intense theoretical effort, aim at unravelling the phase diagram and the thermodynamic properties of QCD \cite{Fukushima:2010bq,Mohanty:2011nm}. 

First-principle calculations based on numerical lattice simulations have demonstrated a crossover between a mostly confined to a mostly deconfined phase as the temperature $T$ is increased \cite{Borsanyi:2013bia}. Such calculations are, however, restricted to small baryon chemical potential $\mu_B$ due to the infamous sign problem that prevents the use of standard Monte Carlo algorithms \cite{deForcrand:2010ys}. A major open question of both theoretical and experimental research programs concerns the possible existence of a first order transition line in the $(T,\mu_B)$ plane, ending at a critical point \cite{Stephanov:2004wx,Schaefer:2007pw}. Continuum approaches can avoid the strong sign problem of the lattice but necessarily rely on approximations or model building that must be tested in situations where lattice results are under control.

In this context, it is of interest to study QCD--like theories, by varying parameters such as the number of colors {and flavors ($N_f$)}, the quark masses, etc.  Investigating the phase structure of the theory in such an extended parameter space, beyond its own theoretical interest, may contribute to our understanding of the phase diagram of the physical theory. An example is the investigation of the Columbia plot, which aims at describing the phase structure of the SU($3$) color theory with two degenerate quark flavors $u$ and $d$ and one strange quark $s$, as a function of the quark masses $M_u=M_d$ and $M_s$  \cite{Pisarski:1983ms,deForcrand:2006pv,Resch:2017vjs}.

The most commonly accepted scenario at zero chemical potential and in the limit of vanishing quark masses is that of a first order transition governed by the restoration of chiral symmetry at large temperatures \cite{Pisarski:1983ms}. This transition weakens as one increases the quark masses and it eventually becomes second order for critical values of the latter. The physical point lies beyond this critical line in the Columbia plot, where the transition is a crossover. A question is, therefore, to study how this critical line evolves with nonzero chemical potential \cite{deForcrand:2006pv}. 

Also of interest is the limit of infinite quark masses, corresponding to the pure Yang-Mills theory. For three colors, the latter presents a first order confinement-deconfinement phase transition in temperature \cite{Svetitsky:1985ye}. Again, this transition weakens for large but finite quark masses and eventually becomes second order for critical values of the quark masses {}\cite{Saito:2011fs}. Contrarily to the light quark case, the heavy quark system can be simulated also at nonzero chemical potential because the sign problem can be evaded by using a large mass expansion \cite{Fromm:2011qi}. In that case, the critical line is known to shrink toward larger quark masses as the chemical potential increases. 

Not only does QCD with heavy quarks provide a framework where one can explicitly study the role of the chemical potential, but it also yields a stringent test case for approximate continuum approaches based either on first principle calculations or on models of the QCD dynamics. Existing calculations include truncations of the Dyson-Schwinger equations (DSE) or functional renormalization group techniques \cite{Fischer:2014vxa}, {one- and two-loop} perturbative calculations in the Curci-Ferrari (CF) model (a massive extension of the Faddeev-Popov (FP) Lagrangian in the Landau gauge) \cite{Reinosa:2015oua,Maelger:2017amh}, and effective models for the Polyakov loop, the order parameter of the confinement-deconfinement transition \cite{Schaefer:2007pw,Pisarski:2000eq,Kashiwa:2012wa}. 

{In the present article, we discuss generic features of the phase diagram of QCD with heavy quarks. It has been observed, both in lattice simulations \cite{Saito:2011fs,Fromm:2011qi} and in matrix models \cite{Kashiwa:2012wa,Kashiwa:2013rm}, that the critical temperature is approximately independent of the quark content and that the flavor dependence of the critical quark masses (which defines the critical line/surface in the Columbia plot) is essentially independent of the details of the gluon dynamics. We recall how the latter comes about for the wide class of models where the quarks are treated at the one-loop level and we explicitly show, in that case, that both the critical temperature and the critical values of the order parameters are constant along the critical line. We further show that these observations extend to finite chemical potential.}

We illustrate these features on various results from the literature, such as the CF model at one-loop \cite{Reinosa:2015oua} and the matrix model of Ref.~\cite{Kashiwa:2012wa}. We also provide results for a model of the glue dynamics recently proposed in Refs.~\cite{Canfora:2015yia,Kroff:2018ncl}, based on the Gribov-Zwanziger (GZ) quantization scheme
\cite{Gribov77}. We {further} show that our generic findings for the critical line extend beyond one-loop as exemplified by the results of the CF model at two-loop order \cite{Maelger:2017amh} and by the DSE results of Ref.~\cite{Fischer:2014vxa}. {Regulator and renormalization effects complicate the comparison between the various continuum and lattice results. We fabricate an observable which suppresses these differences and which exhibits a simple $N_f$ dependence that is satisfied by all existing continuum and lattice results in the large mass limit.  This predicts the critical quark masses for increasing $N_f$---provided confinement is not lost---in all these approaches.}

Let us first consider the case of a vanishing quark chemical potential $\smash{\mu=0}$ and $N_f$ degenerate heavy quark flavors of mass $M$ for simplicity. Descriptions of the QCD dynamics where quarks are included at the one-loop level yield the potential
\begin{equation}\label{eq1}
\beta^4V(\ell,\beta,M)=v_{\rm glue}(\ell,\beta)-2N_f\,f(\beta M)\,\ell\,,
\end{equation}
with $\smash{f(x)=(3x^2/\pi^2)K_2(x)}$, where $K_2(x)$ is the modified Bessel function of the second kind and $\beta^{-1}$ is the temperature. {The Polyakov loop $\ell$ is related to the free energy $F_q$ of a static quark in the thermal bath of gluons as $\ell\propto\exp(-\beta F_q)$ and vanishes identically in the confined phase, where $\smash{F_q=\infty}$}. Here, we assume that the potential $v_{\rm glue}$ is confining, that is, it admits a minimum at $\ell=0$ at zero temperature.

For each value of $N_f$, the critical values for $\ell$, $\beta$ and $M$ on the upper boundary line of the Columbia plot are determined by solving the system of equations ${\partial_\ell V}={\partial_\ell^2 V}={\partial_\ell^3 V}=0$, {that is,}
\begin{align}\label{eq:f}
 {\partial_\ell v_{\rm glue}}&=2N_f f(\beta M)\,,\\
\label{eq:ff}
{\partial_\ell^2 v_{\rm glue}}&={\partial_\ell^3 v_{\rm glue}}=0\,.
\end{align}
The critical values for $\ell$ and $\beta$ are determined from the two equations in (\ref{eq:ff}), which only involve $v_{\rm glue}$, and are, thus completely blind to the quark content of the theory [More precisely, corrections to this statement are suppressed by $\exp(-\beta M)$]. {This explains the fact that the critical temperature $\beta_c^{-1}$ is essentially constant along the critical line in the Columbia plot {\cite{Kashiwa:2012wa,Saito:2011fs,Fromm:2011qi}. The critical value $(\beta M)_c\equiv R_{N_f}$} can be determined from \Eqn{eq:f} for each value of $N_f$. {Since the left-hand side} does not depend on $N_f$, the flavor dependence of $R_{N_f}$ {is determined by the relation \cite{Kashiwa:2012wa}
\begin{equation}\label{eq:3}
N_f f(R_{N_f})=N_f'f(R_{N_f'})\,.
\end{equation}
}
We deduce, in particular, that, even though the critical values $\ell_c$, $\beta_c$ and $R_{N_f}$ depend on { $v_{\rm glue}$, \Eqn{eq:3} is model independent at one-loop order. }

The above discussion easily generalizes to the case of nondegenerate quark masses, where one concludes that the equation determining the critical surface in the space of quark masses is universal, {\it i.e.}, independent of $v_{\rm glue}$. For instance, the critical line in the plane $\smash{(M_{u}=M_d,M_s)}$ of the Columbia plot for $2+1$ (heavy) flavors  is {determined by}
\begin{equation}\label{eq:critline}
2f(\beta M_{u})+f(\beta M_s)=3f(R_3)\,,
\end{equation}
where only the number on the right-hand side, which determines the {\em location} of the critical line, depends on $v_{\rm glue}$, through \Eqn{eq:f}. We can use, alternatively, $3f(R_3)=2f(R_2)=f(R_1)$. This trivially generalizes to more flavors.

The same strategy applies to the case of nonzero {quark chemical potential $\mu=\mu_B/3$}. The effective potential now depends separately on the Polyakov loops $\ell$ and $\bar\ell$ corresponding to quarks and antiquarks, respectively, and reads
\begin{equation}
\beta^4V=v_{\rm glue}(\ell,\bar\ell,\beta)-N_f f(\beta M)(e^{-\beta\mu}\ell+e^{\beta\mu}\bar\ell)\,.
\end{equation}
{For given values of $N_f$ and $\mu$}, the critical point $(\ell_c,\bar\ell_c,\beta_c,M_c)$ is determined by the set of equations\footnote{ For $\mu\in\mathds{R}$, the physical values of $\ell$ and $\bar\ell$ are real. Therefore, it is enough to consider $V(\ell,\bar\ell)$ with real and independent variables. For $\mu\in i\mathds{R}$, one can conveniently work instead with $\bar\ell=\ell^*$ \cite{Dumitru:2005ng}.}
\begin{align}
&{\partial_\ell V}=\partial_{\bar\ell} V=0\,,\label{eq:cond1}\\
 &{\partial_\ell^2 V}{\partial_{\bar\ell}^2 V}-\left({\partial_\ell\partial_{\bar\ell}V}\right)^2=\left(a\partial_\ell-b\partial_{\bar\ell}\right)^3\!V=0\,,\label{eq:cond2}
\end{align}
with $\smash{a=\partial_{\bar\ell}^2V|_c}$ and $\smash{b=\partial_\ell\partial_{\bar\ell}V|_c}$. The first two equations rewrite
\begin{equation}\label{eq:6}
 N_f f(\beta M)=e^{\beta\mu}\,{\partial_\ell v_{\rm glue}}=e^{-\beta\mu}\,\partial_{\bar\ell} v_{\rm glue}\,.
\end{equation}
As in the previous case, the last two equations in \eqn{eq:cond2}, involving two or more $\ell$ and $\bar\ell$ derivatives, only concern $v_{\rm glue}$ and define two functions $\ell(\beta)$ and $\bar\ell(\beta)$ independent of both $N_f$ and $\mu$. Together with these, the ratio of the two equations \eqn{eq:6},
\beq\label{eq:28}
e^{-2\beta\mu} & = & \partial_\ell v_{\rm glue}/\partial_{\bar\ell} v_{\rm glue}\,,
\eeq
determines an $N_f$-independent value for $\beta_c(\mu)$ -- or, more directly, $\mu_c(\beta)$ -- and in turn $N_f$-independent critical values $\smash{\ell_{c}(\mu)=\ell(\beta_c(\mu))}$ and $\smash{\bar\ell_{c}(\mu)=\bar\ell(\beta_c(\mu))}$. As before, the only source of flavor dependence comes from the prefactor of the function $f$ in \Eqn{eq:6}. Taking the ratio of any of these for two different values of $N_f$, we conclude that \Eqn{eq:3} still holds, for any value of $\mu$.

Again, this can be generalized to the case of nondegenerate quark masses and we conclude, following the same {lines}, that the critical values $\ell_c(\mu)$, $\bar\ell_c(\mu)$, and $\beta_c(\mu)$ are constant along the whole critical surface and that the {flavor dependence of the latter is insensitive to $v_{\rm glue}$ {and $\mu$}}. For instance, in the $2+1$ flavor case, the critical line is defined by \Eqn{eq:critline}. Only the actual {\em location} of the critical line depends on $v_{\rm glue}$ and $\mu$, through the determination of $R_3$ {in \Eqn{eq:critline}}. This is demonstrated in \Fig{fig:Columbia}. 

Finally, \Eqn{eq:6} gives the $\mu$ dependence of $R_{N_f}$ in terms of the functions $\beta_c(\mu)$, $\smash{\ell_{c}(\mu)}$ and $\smash{\bar\ell_{c}(\mu)}$, defined above. We also emphasize that the above discussion is valid for both real and imaginary chemical potential. In this respect, it is interesting to mention that the phase diagram at imaginary chemical potential features a tricritical point, which results in a particular scaling of both $\beta_c$ and $R_{N_f}$ when approaching $\smash{\beta\mu=i\pi/3}$ \cite{deForcrand:2002hgr}. The latter can be analytically continued to real $\mu$ and is actually well reproduced by lattice data and by various continuum approaches. The existence of a tricritical point is intimately related to the Roberge-Weiss symmetry of the potential for $\smash{\beta\mu=i\pi/3}$, which reads $\smash{V(\ell,\bar\ell)=V(e^{i2\pi/3}\bar\ell,e^{-i2\pi/3}\ell)}$. To see this, it is convenient to change to the real variables $x$ and $y$ defined by $\smash{\ell=e^{i\pi/3}(x-iy)}$ and $\smash{\bar\ell=e^{-i\pi/3}(x+iy)}$, in terms of which the Roberge-Weiss symmetry rewrites $\smash{y\to -y}$. Defining $x(y)$ from the condition $\smash{\partial_x V=0}$ and evaluating the potential along this {line}, one obtains a reduced potential $V_r(y)$ with $\mathds{Z}_2$ symmetry $\smash{y\to -y}$. The tricritical point corresponds to the cancellation of both the second and the fourth derivatives of $V_r(y)$ at $y=0$. Below, we determine the tricritical point both by following the boundary line of the Columbia plot as $\beta\mu$ approaches $i\pi/3$ or directly from the reduced potential at $\beta\mu=i\pi/3$.

It is interesting to test the above considerations on various one-loop approaches available in the literature. These include the matrix model of Ref.~\cite{Kashiwa:2012wa}, the CF model at one-loop order \cite{Reinosa:2015oua}, as well as the GZ--type model of Refs.~\cite{Canfora:2015yia,Kroff:2018ncl}. In this latter case, no results are available for the Columbia plot. We produce them here after briefly reviewing the approach. 

\begin{figure}
	\centering
	\includegraphics[width=0.45\textwidth]{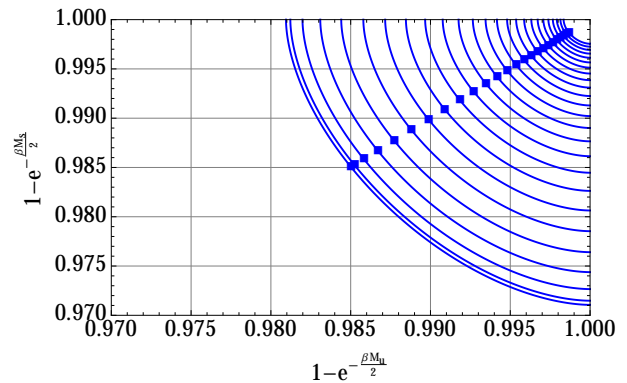}
	\caption{\label{Columbia}Top right corner of the Columbia plot. The squares represent the $\smash{N_f=3}$ model-dependent initial values for $R_{N_f}$, with $\beta\mu\approx 0.2\,n$ and $n\in\mathds{N}$, here obtained from GZ1. The curves are determined uniquely from \Eqn{eq:critline} and lie on top of those obtained from the numerical solution of the model.}
	\label{fig:Columbia}
\end{figure}

 The model for the glue potential proposed in Refs.~\cite{Canfora:2015yia,Kroff:2018ncl} is inspired from the GZ quantization in the vacuum, where the restriction of the path integral to the first Gribov region is implemented by means of a suitably adjusted Gribov parameter.  The extension of this approach to finite temperature proposed in \cite{Canfora:2015yia,Kroff:2018ncl} yields a glue potential in the form $\smash{v_{\rm glue}(\ell,\bar\ell,\beta)=\beta^4W(\ell,\bar\ell,\beta,\{\gamma_\kappa\})}$, where the Gribov parameters $\gamma_\kappa$ can be, {\it a priori} different for each color mode $\kappa$ and are functions of $\ell$, $\bar\ell$, and $T$, determined from the gap equations $\partial_{\gamma_\kappa} W=0$. The function $W$ has been computed to one-loop order in these references, where it was expressed in terms of a background gauge field $\beta \bar A_\mu(x)=\delta_{\mu0}r_jt^j$, with $t^j$ the generators of the Cartan subalgebra of the gauge group. For SU($3$), these are the two diagonal Gell-Mann matrices $\lambda^{(3)}/2$ and $\lambda^{(8)}/2$. At this order of approximation, the background field is related to the Polyakov loops as 
\begin{equation}
\ell(r)=\bar\ell(-r)= \frac{1}{3}\left[e^{-i\frac{r_8}{\sqrt{3}}}+2e^{i\frac{r_8}{2\sqrt{3}}}\cos(r_3/2)\right].
\end{equation}

The color modes $\kappa$ correspond either to one of the Cartan directions $j$ or to any of the roots $\alpha$ of the gauge algebra. In Ref.~\cite{Kroff:2018ncl}, three scenarios for the color dependence of the Gribov parameters have been proposed, all compatible with the symmetries of the problem:

\vspace{.2cm}
\hglue1mm {\bf GZ1:} $\gamma_\kappa=\gamma\,,\,\,\,\forall \kappa$\,.

\hglue1mm {\bf GZ2:} $\gamma_{\kappa}=\gamma_1$ if $\kappa=j$ and $\gamma_\alpha=\gamma_2\,,\,\,\,\forall \alpha$\,.

\hglue1mm {\bf GZ3:} $\gamma_{\kappa}=\gamma_1$ if $\kappa=j$ and $\gamma_\alpha=\gamma_{-\alpha}$\,.
\vspace{.2cm}

\noindent
We stress that, when computing the derivatives of the glue potential in Eqs. \eqn{eq:cond1} and \eqn{eq:cond2}, one needs to take into account the $\ell$ and $\bar\ell$ dependence of the Gribov parameters. Taking derivatives of the gap equation $\smash{\partial_{\gamma}W=0}$, one gets, for example in the case GZ1, $\partial_\ell \gamma=-\partial_{\ell,\gamma}^2W/\partial_{\gamma,\gamma}^2 W$ {and similarly for other derivatives.}

\begin{table}[h!]
  \centering
	\begin{tabular}{|c | c c c || c  c || c |} 
		\hline
		 $\mu=0$ & $\,\,\,R_1\,\,\,$ & $\,\,\,R_2\,\,\,$  &$\,\,\,R_3\,\,\,$ & $R_2/R_1$ & $R_3/R_2$ & $Y_3$\\ [0.5ex] 
		\hline\hline
		Lattice \cite{Fromm:2011qi} & 7.23 & 7.92 &  8.33 & 1.10 & 1.05 & 1.59\\
		\hline\hline
		GZ1 & 7.09 & 7.92 & 8.40 & 1.12 & 1.06 & {1.58} \\
		\hline
		GZ2 & 9.45 & 10.25 & 10.72 & 1.08 & 1.05 & {1.58} \\
		\hline
		GZ3 &  $\cdots$ & 1.33 & 2.12 & $\cdots$ & 1.59 & $\cdots$ \\
		\hline
		GZ0 &  4.66 &  5.56 & 6.07 & 1.20 &  1.09 &  1.59 \\
		\hline
		Matrix \cite{Kashiwa:2012wa} & 8.04 & 8.85 & 9.33 & 1.10 & 1.05 & 1.59\\
		\hline
		CF1 \cite{Reinosa:2015oua} & 6.74 & 7.59 & 8.07 &1.12 & 1.06 & {1.58}\\ 
		\hline\hline
		CF2 \cite{Maelger:2017amh} & 7.53 & 8.40 & 8.90 & { 1.11} & 1.06 &  {1.57}\\
		\hline
		DSE \cite{Fischer:2014vxa} & 1.42 & 1.83 & 2.04 & 1.29 & 1.11 & 1.51\\
		\hline
	\end{tabular}
    	\caption{$R_{N_f}$ for $N_f=1$, $2$, and $3$ degenerate quark flavors, as computed in various approaches. CF1 and CF2 refer to the one- and two-loop results within the CF model. The last two lines of the table gather results beyond one-loop order.}
	\label{ZeroCritics}
\end{table}

We  compute the values of $R_{N_f}$ for various $N_f$ in each scenario and compare them to lattice results. At the order of approximation considered here, these quantities do not depend on the value of the Gribov parameter at zero temperature and are, therefore, a stringent test for the various scenarios. Our results are gathered in Table~\ref{ZeroCritics} and compared to other approaches. We observe that, among the various one-loop approaches considered here, the degenerate GZ approach  (GZ1) gives the best results as compared to the lattice values. Neglecting the background dependence of the $\gamma_\kappa$'s yields the values referred to as GZ0. We also note that the nondegenerate case (GZ3) is completely discarded : For $N_f=1$, we find a first order phase transition irrespectively of the value of the quark mass and the values for $R_2$ and $R_3$ are way too small.

One easily checks that the universal equation (\ref{eq:3}) is accurately satisfied, given the precision, for all one-loop results in Table~\ref{ZeroCritics}. In contrast, the lattice values do not satisfy the scaling \eqn{eq:3} well. As we discuss below, they follow a somewhat different scaling, where the function $f(x)$ is replaced by a simple exponential.

In Table~\ref{ZeroCritics}, we also quote the results from two calculations beyond one-loop order, namely, the two-loop perturbative calculation in the CF model of Ref.~\cite{Maelger:2017amh} and the DSE results of Ref.~\cite{Fischer:2014vxa}. In this case, the direct comparison of critical masses is complicated by nontrivial mass renormalization beyond leading order. As argued in Ref.~\cite{Maelger:2017amh}, this can be partially absorbed in the ratios $R_{N_f}/R_{N_f'}$. In particular, these ratios for DSE agree well with the other continuum approaches despite the low values of $R_{N_f}$. Note that the same argumentation does not save the scenario GZ3, first, because there is no reason to expect such mass renormalization effect at one-loop order and, second, because the ratio $R_3/R_2$ is bad anyway as compared to all other continuum approaches.

In the same vein, as discussed in Ref.~\cite{Maelger:2017amh}, the lattice also brings additive mass renormalizations because it explicitly breaks chiral symmetry. {Such effects are suppressed by taking ratios of differences, {\it e.g.},}
\begin{equation}
Y_{N_f}=\frac{R_{N_f}-R_{1}}{R_{2}-R_{1}}.
\end{equation}
In Table~\ref{ZeroCritics}, we quote the results for $Y_3$, which is surprisingly stable for all the approaches (continuum and lattice). As far as continuum models of the form \eqn{eq1} are concerned, this {is a simple consequence of the scaling law \eqn{eq:3} and the large values of $R_{N_f}$. Using the asymptotic expansion of the Bessel function, one easily checks that}
\beq\label{eq:20}
Y_{N_f}\approx\frac{\ln N_f}{\ln 2},
\eeq
up to relative corrections of ${\cal O}(R_1^{-2})$. 
{This yields $Y_3\approx1.58$, in excellent agreement with our results in Table~\ref{ZeroCritics}.} 

It is remarkable that the results for $Y_3$ for all approaches beyond one-loop agree well with the prediction \eqn{eq:20}. We can understand the lattice value from the $N_f$ scaling observed in Ref.~\cite{Fromm:2011qi}
\begin{equation}\label{eq:latticescaling}
N_f e^{-R_{N_f}}\cosh (\beta_c\mu)\approx {0.00075}\,,
\end{equation}
valid for all values of $\mu$. This yields the scaling law $\exp(R_{N_f'}-R_{N_f})=N_f'/N_f$, from which one immediately obtains \Eqn{eq:20} using the fact that the lattice value of $\beta_c$ is {found insensitive to $N_f$.} {Similarly, at leading order in the hopping ({\it i.e.}, large quark mass) expansion, the lattice results satisfy the law \eqn{eq:critline} with $f(x)\to e^{-x}$ \cite{Saito:2011fs}.}

We conclude that the scaling law \eqn{eq:20} is quite robust and universal (independent of the actual gluon dynamics) and can thus be used to predict the values of $R_{N_f}$ for $N_f\ge3$, given $R_2$ and $R_1$ in the various approaches. We have checked that these predictions describe well the actual results in the one-loop models GZ1, GZ2, and CF1, as expected. We {have also verified} that the same holds for the two-loop values of $R_{N_f}$ in CF2.

\begin{table}[h!]

\centering
\begin{tabular}{|c | c c c || c c || c |}
\hline
$\beta\mu=i\pi/3$  & $\,\,\, R_1\,\,\,$ & $\,\,\, R_2\,\,\,$  &$\,\,\, R_3\,\,\,$ & $R_2/R_1$ & $R_3/R_2$ & $Y_3$ \\ [0.5ex]
\hline\hline
Lattice \cite{Fromm:2011qi} & 5.56 & 6.25 & 6.66 & 1.12 & 1.07 & 1.59 \\
\hline\hline
GZ1 & 5.02 & 5.91 & 6.42 & 1.18 & 1.09  & 1.57 \\
\hline
GZ2 & 7.51 & 8.34 & 8.82 & 1.11 & 1.06 & 1.58 \\
\hline
Matrix {\cite{Kashiwa:2013rm} } & 5.00 & 5.90 & 6.40 & 1.18 & 1.08 & 1.56 \\
\hline
CF1 \cite{Reinosa:2015oua} & 4.72 & 5.63 & 6.14 & 1.19 & 1.09 & 1.57 \\
\hline\hline
CF2 \cite{Maelger:2017amh} & 5.47 & 6.41 & 6.94 & 1.17 & 1.08 & 1.57\\
\hline
DSE \cite{Fischer:2014vxa} & 0.41 & 0.85 & 1.11 & 2.07 & 1.31 & 1.59  \\
\hline
\end{tabular}
\caption{$R_{N_f}$ at $\beta\mu=i\pi/3$, for $N_f=1$, $2$ and $3$ degenerate quark flavors, as computed in various approaches.}
\label{ImagCritics}
\end{table}

The results for an imaginary chemical potential, and more precisely along the tricritical line at $\smash{\beta\mu=i\pi/3}$, are gathered in Table~\ref{ImagCritics}. We verify that the law (\ref{eq:3}) and its consequences as described above are well verified by the values of $R_{N_f}$  for one-loop models. Again, the GZ1 scenario seems to be giving the best values to date, although they are in this case pretty close to the ones from the matrix model of {Ref.~\cite{Kashiwa:2013rm}}. 

As anticipated, the values of $Y_3$ from all approaches agree well with the prediction \eqn{eq:20}. This is to be expected from the above discussion for both one-loop models and Lattice results. The DSE value is strikingly good despite the low values of $R_{N_f}$ and the not so good values of the ratios $R_{N_f}/R_{N_f'}$. This supports the conjecture of the universal character of $Y_{N_f}$.

Acknowledgement: We thank Daniel Kroff for related work on the Gribov-Zwanziger model.


\begin{thebibliography}{1}
 
\bibitem{Weissenborn:2011qu}
  S.~Weissenborn, I.~Sagert, G.~Pagliara, M.~Hempel and J.~Schaffner-Bielich,
  Astrophys.\ J.\  {\bf 740} (2011) L14;
  S.~Bors\'anyi {\it et al.},
  Nature (London) {\bf 539} (2016)  69.
 
  
\bibitem{Fukushima:2010bq}
  K.~Fukushima and T.~Hatsuda,
  Rept.\ Prog.\ Phys.\  {\bf 74} (2011) 014001.
  H.~T.~Ding, F.~Karsch and S.~Mukherjee,
  Int.\ J.\ Mod.\ Phys.\ E {\bf 24} (2015) no.10,  1530007
  
\bibitem{Mohanty:2011nm}
  B.~Mohanty (STAR Collaboration),
  J.\ Phys.\ G {\bf 38} (2011) 124023;
  T.~Ablyazimov {\it et al.} [CBM Collaboration],
  Eur.\ Phys.\ J.\ A {\bf 53} (2017)  60;
  P.~Senger,
  Eur.\ Phys.\ J.\ A {\bf 52} (2016) 217;
  T.~Sakaguchi [J-PARC-HI Collaboration],
  Nucl.\ Phys.\ A {\bf 967} (2017) 896.

  
\bibitem{Borsanyi:2013bia}
  S.~Bors\'anyi, Z.~Fodor, C.~Hoelbling, S.~D.~Katz, S.~Krieg, and K.~K.~Szabo,
  Phys.\ Lett.\ B {\bf 730} (2014) 99.

\bibitem{deForcrand:2010ys}
  P.~de Forcrand,
  PoS LAT {\bf 2009} (2009) 010;
  O.~Philipsen,
  arXiv:1009.4089 [hep-lat].


\bibitem{Stephanov:2004wx}
  M.~A.~Stephanov,
  Prog.\ Theor.\ Phys.\ Suppl.\  {\bf 153} (2004) 139
   [Int.\ J.\ Mod.\ Phys.\ A {\bf 20} (2005) 4387];
  P.~de Forcrand, S.~Kim, and O.~Philipsen,
  PoS LAT {\bf 2007} (2007) 178;
  C.~S.~Fischer, J.~Luecker and J.~A.~Mueller,
  Phys.\ Lett.\ B {\bf 702} (2011) 438.

\bibitem{Schaefer:2007pw}
  B.~J.~Schaefer, J.~M.~Pawlowski and J.~Wambach,
  Phys.\ Rev.\ D {\bf 76} (2007) 074023.

\bibitem{Pisarski:1983ms}
  R.~D.~Pisarski and F.~Wilczek,
  Phys.\ Rev.\ D {\bf 29} (1984) 338.

\bibitem{deForcrand:2006pv}
  P.~de Forcrand and O.~Philipsen,
  JHEP {\bf 0701} (2007) 077;

\bibitem{Resch:2017vjs}
  S.~Resch, F.~Rennecke and B.~J.~Schaefer,
  arXiv:1712.07961 [hep-ph].
    
\bibitem{Svetitsky:1985ye}
  B.~Svetitsky,
  Phys.\ Rep.\  {\bf 132} (1986) 1.

 
\bibitem{Saito:2011fs}
  H.~Saito {\it et al.} [WHOT-QCD Collaboration],
  Phys.\ Rev.\ D {\bf 84} (2011) 054502;
   Erratum: Phys.\ Rev.\ D {\bf 85} (2012) 079902.
  

  
\bibitem{Fromm:2011qi}
  M.~Fromm, J.~Langelage, S.~Lottini and O.~Philipsen,
  JHEP {\bf 1201} (2012) 042.
   
\bibitem{Fischer:2014vxa}
  C.~S.~Fischer, J.~Luecker and J.~M.~Pawlowski,
  Phys.\ Rev.\ D {\bf 91} (2015)  014024;
  J.~Braun, L.~M.~Haas, F.~Marhauser and J.~M.~Pawlowski,
  Phys.\ Rev.\ Lett.\  {\bf 106}, 022002 (2011).
  
\bibitem{Reinosa:2015oua} 
  U.~Reinosa, J.~Serreau and M.~Tissier,
  Phys.\ Rev.\ D {\bf 92}, 025021 (2015).
 
\bibitem{Maelger:2017amh} 
  J.~Maelger, U.~Reinosa and J.~Serreau,
  Phys.\ Rev.\ D {\bf 97} 074027 (2018).
  
\bibitem{Pisarski:2000eq}
  R.~D.~Pisarski,
  Phys.\ Rev.\ D {\bf 62} (2000) 111501;
  A.~Dumitru, D.~Roder and J.~Ruppert,
  Phys.\ Rev.\ D {\bf 70} (2004) 074001;
  P.~M.~Lo, B.~Friman and K.~Redlich,
  Phys.\ Rev.\ D {\bf 90} (2014)  074035.

\bibitem{Kashiwa:2012wa}
  K.~Kashiwa, R.~D.~Pisarski and V.~V.~Skokov,
  Phys.\ Rev.\ D {\bf 85} (2012) 114029.

\bibitem{Kashiwa:2013rm}
  K.~Kashiwa and R.~D.~Pisarski,
  Phys.\ Rev.\ D {\bf 87} (2013) 096009.


\bibitem{Canfora:2015yia} 
  F.~E.~Canfora, D.~Dudal, I.~F.~Justo, P.~Pais, L.~Rosa and D.~Vercauteren,
  Eur.\ Phys.\ J.\ C {\bf 75}, no. 7, 326 (2015).
  
\bibitem{Kroff:2018ncl} 
  D.~Kroff and U.~Reinosa,
  Phys. Rev. D98 (2018) 034029.
 
\bibitem{Gribov77}
  V.~N.~Gribov,
  Nucl.\ Phys.\  B {\bf 139} (1978) 1;
  D.~Zwanziger,
  Nucl.\ Phys.\  {\bf B323}, 513 (1989);
  Nucl.\ Phys.\  {\bf B399}, 477 (1993);
  N.~Vandersickel and D.~Zwanziger,
  Phys.\ Rep.\  {\bf 520}, 175 (2012).
    
\bibitem{Dumitru:2005ng} 
  A.~Dumitru, R.~D.~Pisarski and D.~Zschiesche,
  Phys.\ Rev.\ D {\bf 72}, 065008 (2005).
  
\bibitem{deForcrand:2002hgr}
  P.~de Forcrand and O.~Philipsen,
  Nucl.\ Phys.\ B {\bf 642} (2002) 290;
  Phys.\ Rev.\ Lett.\  {\bf 105} (2010) 152001.

  
\end{thebibliography}
  \end{document}